\documentclass[reprint, amsmath,amssymb, aps, pra, nofootinbib]{revtex4-2}

\usepackage[T2A]{fontenc}                   
\usepackage[utf8]{inputenc}                 
\usepackage[russian,english]{babel}         
\usepackage[margin=1.8cm]{geometry}         

\usepackage[unicode, pdftex]{hyperref}      

\usepackage{amsthm}                         
\usepackage{hypcap}                         
\usepackage{caption}                        
\usepackage{fancyhdr}                       
\usepackage{wrapfig}                        

\usepackage{ragged2e} % provides \justifying

\usepackage{amsmath}                        
\usepackage{amssymb,textcomp, esvect,esint} 
\usepackage{amsfonts}                       
\usepackage{mathrsfs}                       

\usepackage{abraces}                        
\usepackage{pifont}                         
\usepackage{cancel}                         
\usepackage{esvect}                         

\usepackage{graphicx}                       
\usepackage{indentfirst}                    
\usepackage{xcolor}                         
\usepackage{enumitem}                       

\usepackage{booktabs}                       
\usepackage{multirow}                       
\usepackage{array}

\usepackage{bbm}
\usepackage{newfloat}
\usepackage{orcidlink}

\definecolor{ugrey}{HTML}{666666}
\definecolor{ublue}{HTML}{08088A}
\newcommand{\reccall}[1]{\setlength{\fboxsep}{1pt}{\colorbox{blue!12}{$\displaystyle #1$}}}
\newcommand{\inverseoftwo}[1]{\setlength{\fboxsep}{1pt}{\colorbox{red!8}{$\displaystyle #1$}}}

\renewcommand{\leq}{\leqslant}
\renewcommand{\geq}{\geqslant}

\newcommand{\vc}[1]{\boldsymbol{#1}}

\newcommand{\T}{^{\textnormal{T}}}

\newcommand{\stmm}[1]{\mathtt{#1}}

\newcommand{\red}[1]{\textcolor{red}{#1}}
\newcommand{\blue}[1]{\textcolor{ublue}{#1}}
\newcommand{\grey}[1]{\textcolor{ugrey}{#1}}

\DeclareDocumentCommand{\bk}{m o m}{
    \IfNoValueTF{#2}{\langle #1 | #3 \rangle}{\langle #1 | #2 | #3 \rangle}
}
\DeclareDocumentCommand{\kb}{m o m}{
    \IfNoValueTF{#2}{| #1 \rangle \langle #3 |}{| #1 \rangle #2 \langle #3 |}
}

\newcommand{\F}{\mathbb{F}}
\newcommand{\Z}{\mathbb{Z}}

\newcommand{\Q}{\mathbb{Q}}

\newcommand{\usubsec}[1]{}

\newcommand{\tabiemph}[1]{\textbf{#1}}

\usepackage{titlesec}

\makeatletter
\renewcommand\thesection{\arabic{section}}
\makeatother

\titleformat{\section}[block]{\normalfont\normalsize\scshape\centering}{\thesection.}{0.75em}{}
\titlespacing*{\section}{0pt}{\baselineskip}{0.5\baselineskip}
\titlespacing*{\subsection}{0pt}{\baselineskip}{0.25\baselineskip}

\usepackage{fancyhdr}
\fancypagestyle{plain}{%
  \fancyhf{}
  \fancyfoot[C]{\thepage}

}

\begin{document}

\setlength{\abovedisplayskip}{3pt}
\setlength{\abovedisplayshortskip}{3pt}
\setlength{\belowdisplayskip}{3pt}
\setlength{\belowdisplayshortskip}{3pt}
\setlength{\headheight}{13pt}
\setlength{\parskip}{0pt}

\title{Faster Algorithms for Structured Matrix Multiplication via Flip Graph Search}

\author{
	Kirill Khoruzhii\orcidlink{0000-0003-4689-3812}$^{1,*}$,
	Patrick Gelß\orcidlink{0000-0002-3645-9513}$^{1}$,
    Sebastian Pokutta\orcidlink{0000-0001-7365-3000}$^{1,2}$
}
\affiliation{
	$^1$Zuse Institute Berlin, Berlin, Germany\\
    $^2$Technische Universität Berlin, Germany
}

\begin{abstract}
We give explicit low-rank bilinear non-commutative schemes for multiplying structured $n \times n$ matrices with $2 \leq n \leq 5$, which serve as building blocks for recursive algorithms with improved multiplicative factors in asymptotic complexity. Our schemes are discovered over $\F_2$ or $\F_3$ and lifted to $\Z$ or $\Q$. Using a flip graph search over tensor decompositions, we derive schemes for general, upper-triangular, lower-triangular, symmetric, and skew-symmetric inputs, as well as products of a structured matrix with its transpose. These schemes improve asymptotic constants for 13 of 15 structured formats. In particular, we obtain $4 \times 4$ rank-$34$ schemes for both multiplying a general matrix by its transpose and an upper-triangular matrix by a general matrix, improving the asymptotic factor from $8/13$ (0.615) to $22/37$ (0.595). Additionally, using $\F_3$ flip graphs, we discover schemes over $\Q$ that fundamentally require the inverse of $2$, including a $2 \times 2$ symmetric-symmetric multiplication of rank $5$ and a $3 \times 3$ skew-symmetric-general multiplication of rank $14$ (improving upon AlphaTensor's $15$).
\end{abstract}

\maketitle
\thispagestyle{fancy}

\begin{figure*}[t]
    \centering
    \includegraphics{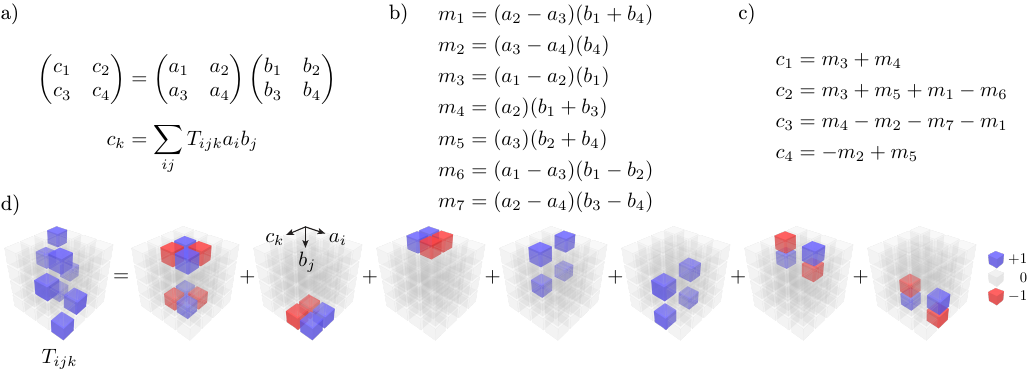}
    \caption{ \justifying
        \textbf{Strassen’s decomposition rank-$7$ for $2\times2$ matrix multiplication.} 
        a) Matrix product written as a bilinear contraction. 
        b) The seven intermediate products. 
        c) Output reconstructed entries. 
        d) Tensor view: $T_{ijk}$ expressed as a sum of seven rank-1 terms; each term corresponds to one intermediate product $m_\ell$ in (b) and encodes how it contributes to each output $c_k$ in (c). 
    }
    \label{fig:gg-222}
\end{figure*}

\begingroup
\renewcommand\thefootnote{\fnsymbol{footnote}}
\footnotetext[1]{khoruzhii@zib.de}
\endgroup

\section{Introduction}

Modern computational workflows rely heavily on matrix multiplication, in particular for matrices with exploitable structure. Expressions such as $XX\T$ appear throughout statistics as covariance or Gram matrices, in optimization algorithms for computing Newton-Schulz updates in modern LLM training methods \cite{rybin_2025_05,vyas_2025_01,jordan_2025_10}, and in linear regression where solutions involve the data covariance matrix. Similarly, triangular matrix multiplication arises naturally when working with factorizations or solving triangular systems, and also in the masking used for causal attention in transformer-based neural networks \cite{rybin_2025_10}. Unlike general matrix multiplication, these structured products offer opportunities for constant-factor improvements even when asymptotic complexity matches the general case. These operations correspond to Level 3 BLAS primitives: SYRK (symmetric rank-k update) computes $AA\T$, TRMM (triangular matrix multiply) handles products involving triangular matrices, and related operations form the computational backbone of numerical linear algebra libraries. Despite their prevalence in applications, structured matrix-matrix products have received far less systematic attention \cite{rybin_2025_05} than general multiplication in the algorithmic discovery literature.

Since Strassen showed the naive $O(n^3)$ algorithm is suboptimal~\cite{strassen_1969_08}, research has followed two main directions. One branch pursues improvements to the asymptotic exponent $\omega$, where recent laser-method refinements~\cite{alman_2025_01} have pushed the bound to $\omega < 2.372$. The other branch focuses on practical algorithms \cite{pan_1982_01,schwartz_2025_08} for fixed small matrix sizes as base, seeking low-rank bilinear decompositions that minimize the number of scalar multiplications for specific formats. Computing or even approximating the minimal tensor rank is NP-hard~\cite{hastad_1990_12}. Constant-factor improvements translate to performance gains in practice, particularly for recursive algorithms where small-base schemes serve as building blocks. For structured products, the asymptotic complexity remains $\Theta(n^\omega)$ as it does for general multiplication, but the multiplicative constants can differ, creating opportunities for specialized algorithms that exploit structure.

Automated methods now efficiently find low-rank tensor decompositions. Numerical optimization~\cite{smirnov_2013_12,kaporin_2024_09}, and deep reinforcement learning approaches such as AlphaTensor~\cite{fawzi_2022_10} have successfully discovered state-of-the-art schemes for general matrix multiplication, though often requiring substantial computational resources. Random walks in flip graphs~\cite{kauers_2023_07} enable discovery of competitive schemes on standard hardware by restricting search to correct decompositions. However, systematic application of these automated methods has focused almost exclusively on general matrix multiplication \cite{arai_2024_03, moosbauer_2023_09, moosbauer_2025_02,kauers_2025_10,heule_2019_08} and polynomial multiplication~\cite{chen_2025_02}. For structured products, prior work consists of representation-theoretic matrix-vector constructions~\cite{ye_2016_06,ye_2018_02} and scattered results for specific formats~\cite{dumas_2021_01,fawzi_2022_10,rybin_2025_05}, but no comprehensive framework exists for discovering and cataloging schemes across the full range of structured matrix-matrix operations.

We systematically explore structured matrix multiplication via flip graph search and catalog schemes for all 15 distinct format combinations (after accounting for symmetries) with base sizes $n \in \{2, 3, 4, 5\}$. We extend this methodology in two key directions. First, we perform searches over $\F_3$ in addition to $\F_2$, enabling discovery of schemes that fundamentally require the inverse of 2 and successfully lift them to $\Q$. Second, we introduce specialized techniques for structured tensors that, in transpose-product formats, increase the fraction of recursive calls. Our approach achieves systematic exploration of the structured multiplication landscape on standard hardware, improving asymptotic complexity factors for 13 of 15 structured formats considered. Notable improvements include computing $AA^T$ and triangular-general multiplication, both corresponding to fundamental Level 3 BLAS primitives. We also discover schemes over $\F_3$ requiring the inverse of 2, including a $2 \times 2$ symmetric-symmetric multiplication of rank 5 and a $3 \times 3$ skew-symmetric-general multiplication of rank 14, improving on ranks 6 and 15, respectively \cite{ye_2018_02,fawzi_2022_10}.

The remainder of this paper is organized as follows. Section~\ref{sec:algorithms} reviews matrix multiplication as tensor decomposition and introduces our notation for structured formats. Section~\ref{sec:stmm} defines structured matrix multiplication formally, establishes the notation $\langle n_1,n_2,n_3 : r\rangle_\stmm{ab}^{(q_\stmm{ab},q_\stmm{ag},q_\stmm{gb})}$ for recursive schemes, and derives the general asymptotic complexity formula. Section~\ref{sec:flipgraph} describes the flip graph construction, proves its connectivity properties, and explains the operations (flips, reductions, plus-transitions) that generate edges. Section~\ref{sec:recursive} details our methodology for identifying recursive schemes through Hensel lifting from $\F_2$ or $\F_3$ to $\Z$ or $\Q$, including our corner-zeroing technique for transpose products. Section~\ref{sec:results} presents our complete results with comparisons to prior work. Section~\ref{sec:discussion} discusses extensions to other bilinear computations, multi-objective optimization challenges, and prospects for directed search on flip graphs.

\begin{figure*}[t]
    \centering
    \includegraphics{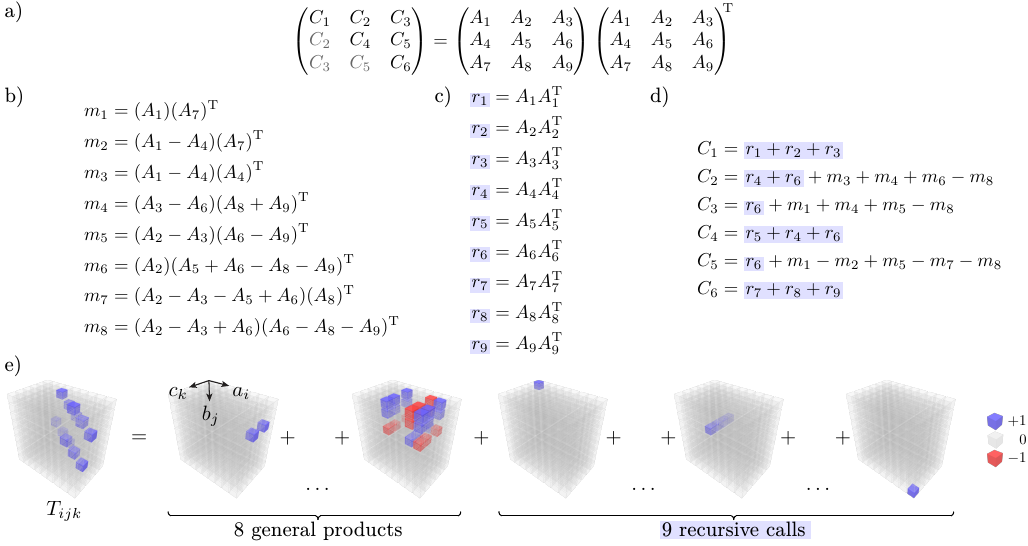}
    \caption{ \justifying
        \textbf{Decomposition rank-$17$ for $3 \times 3$ matrix multiplication $G G\T$.}
        a) Block layout $C=GG\T$ with $G$ partitioned into $3\times3$ blocks $A_1,\ldots,A_9$; by symmetry only six distinct blocks $C_1,\ldots,C_6$ should be calculated.
        b) General products $m_1,\ldots,m_8$.
        c) Nine recursive calls $r_1,\ldots,r_9$ computing the symmetric block products $A_iA_i\T$.
        d) Reconstruction of $C_1,\ldots,C_6$ from the $m$’s and $r$’s.
        e) Tensor view: contraction tensor $T_{ijk}$ expressed as a sum of rank-1 terms.
    }
    \label{fig:gt-333}
\end{figure*}

\section{Algorithms as Tensor Decomposition} \label{sec:algorithms}

Matrix multiplication (Fig.~\ref{fig:gg-222}a) can be written as a bilinear mapping $(\vc{a}, \vc{b}) \mapsto \vc{c}$ with $c_k = \sum_{i,j} T_{ijk}\, a_i b_j$. A canonical polyadic decomposition (CPD) of the tensor
\begin{equation}
    T_{ijk} = \sum_{q=1}^{r} U_{qi} V_{qj} W_{qk}
    \label{eq:cpd}
\end{equation}
provides a recipe to compute $\vc{c}$ using $r$ multiplications:
\begin{equation}
    c_k = \sum_{q=1}^r W_{qk} \big(\textstyle{\sum_i} U_{qi} a_i\big) \big(\textstyle{\sum_j} V_{qj} b_j\big).
    \label{eq:bilinear}
\end{equation}
Here $a_i$ and $b_j$ denote entries of the vectorized input matrices. We regard multiplications by $U$ and $V$ (linear combinations of inputs) as inexpensive compared to the $r$ scalar multiplications between the two parenthesized terms. Algorithmic complexity equals $r$, the number of scalar multiplications.

The coefficients $U_{qi}$, $V_{qj}$, $W_{qk}$ in~\eqref{eq:cpd} can be integers, rationals, or elements of finite fields. We discover schemes over the finite fields $\F_2$ and $\F_3$, then apply Hensel lifting to obtain schemes over $\Z$ or $\Q$ (see Section~\ref{sec:flipgraph}). Some schemes fundamentally require the inverse of 2 and therefore exist only over $\Q$, not over $\Z$. We report schemes with both integer and rational coefficients.

The \emph{tensor rank} of $T_{ijk}$ is the minimal number of terms in a CP decomposition~\eqref{eq:cpd}. In contrast, the \emph{scheme rank} is the number of terms $r$ in a given decomposition. Any $r$-term scheme provides an upper bound for the tensor rank, but computing the exact tensor rank is NP-hard~\cite{hastad_1990_12}. Our goal is to discover schemes with rank as close to the tensor rank as possible, thereby minimizing the number of scalar multiplications required. Throughout, unless stated otherwise, \emph{rank} refers to the scheme rank.

For matrix multiplication schemes, we use the notation~\cite{dumas_2025_06} $\langle n_1, n_2, n_3 : r \rangle$ to denote an algorithm for multiplying $n_1 \times n_2$ by $n_2 \times n_3$ matrices using $r$ scalar multiplications. The corresponding tensor $T_{ijk}$ is determined entirely by the dimensions $(n_1, n_2, n_3)$ and has size $(n_1 n_2, n_2 n_3, n_1 n_3)$, corresponding to the vectorized left input, right input, and output. In this paper we focus on the square case $n_1 = n_2 = n_3$. For example, Strassen's algorithm is denoted $\langle 2, 2, 2 : 7 \rangle$. Fig.~\ref{fig:gg-222} illustrates this scheme in multiple representations: the algebraic formulation (b, c), the tensor decomposition view (d), and the graphical visualization. When applied recursively to $n \times n$ matrices by partitioning them into $2 \times 2$ blocks, Strassen's scheme yields complexity $O(n^{\log_2 7}) \approx O(n^{2.807})$.

The bilinear formulation~\eqref{eq:bilinear} preserves the order of factors $a_i$ and $b_j$, meaning that these entries can be non-commutative objects such as matrices themselves. This property enables recursive application: each scalar multiplication in the scheme can be replaced by a smaller matrix multiplication, and the scheme remains correct. Iterating this process level by level produces algorithms with sub-cubic exponents. Throughout this work, we focus exclusively on the non-commutative case, as it is essential for recursive block matrix multiplication.

\section{Structured Matrix Multiplication} \label{sec:stmm}

\begin{figure*}[t]
    \centering
    \includegraphics{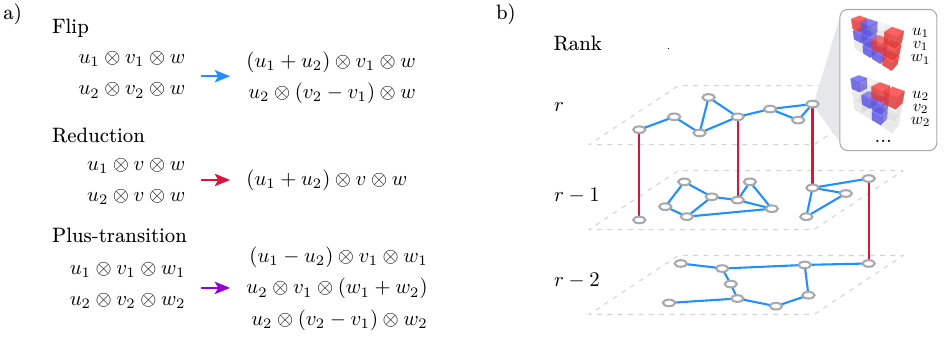}
    \caption{\justifying
        \textbf{Flip graph structure and operations.} 
        a) Three types of transformations between tensor decompositions: \emph{flip} (blue) modifies two rank-1 terms sharing a common factor, preserving the total rank; \emph{reduction} (red) eliminates one term when two rank-1 terms share two common factors; \emph{plus-transition} (purple) combines an inverse reduction with a flip to escape local plateaus. 
        b) The flip graph organizes schemes by rank. Vertices represent correct matrix multiplication schemes (shown as sum of rank-1 tensors $u_i \otimes v_i \otimes w_i$). Horizontal edges (blue) correspond to flips within a fixed rank level. Vertical edges (red) correspond to reductions that decrease rank. Some connected components at rank $r-1$ may have no further reductions, necessitating plus-transitions to continue descent.
    }
    \label{fig:flip-graph}
\end{figure*}

Consider computing $C = A A\T$. Unlike general matrix multiplication, this product exhibits exploitable structure: the result is symmetric ($C = C\T$), so we need only compute its upper triangular part. In naive algorithms, this reduces the workload from $n^3$ scalar multiplications to $\tfrac{1}{2}n(n+1) \times n$ multiplications for $n \times n$ matrices.

In terms of the multiplication tensor, this structure translates to a reduced output dimension. For a scheme $\langle n_1, n_2, n_1 \rangle$ with tensor dimension $(n_1 n_2, n_1 n_2, n_1 n_1)$, we retain only the upper triangular output components, yielding effective dimension $(n_1 n_2, n_1 n_2, n_1(n_1+1)/2)$. Removing output axes often enables schemes with lower rank. Fig.~\ref{fig:gt-333} illustrates this for $n=3$: a rank-17 scheme computes $AA\T$ using only 17 multiplications instead of the naive 18. Crucially, our bilinear formulation preserves the order of factors $a_i$ and $b_j$ in equation~\eqref{eq:bilinear}, making these schemes non-commutative and thus applicable to recursive block matrix multiplication.

In this paper we consider the following matrix structures: \emph{general} ($\stmm{g}$, no constraints), \emph{upper-triangular} ($\stmm{u}$), \emph{lower-triangular} ($\stmm{l}$), \emph{symmetric} ($\stmm{s}$), and \emph{skew-symmetric} ($\stmm{k}$). For products involving a transpose $(AA\T)$ we use the notation $\stmm{t}$ to mark the operation. The result is always symmetric, and we compute only its upper triangular part. For example, $\stmm{gt}$ denotes multiplication of a general matrix $A$ by its transpose to produce $AA\T$.

Skew-symmetric matrices have zero diagonals, but in block algorithms diagonal blocks are skew-symmetric matrices. To handle this, we introduce structure $\stmm{w}$ (skew-symmetric plus diagonal), which allows diagonal blocks to inherit the skew-symmetric structure. Recursive schemes are therefore constructed using $\stmm{w}$, though we report complexity results for the standard $\stmm{k}$ structure (computed via $\stmm{w}$-based schemes).

\usubsec{Asymptotic Complexity}
Let $M_{\stmm{ab}}(n)$ denote the number of scalar multiplications required to compute the product of $n \times n$ matrices with structures $\stmm{a}$ and $\stmm{b}$.
Although structured matrix multiplication requires fewer scalar operations than the general case, the asymptotic complexity remains $O(n^\omega)$. To see this, consider the block concatenation
\begin{equation*}
    X = \begin{bmatrix} A \\ B\T \end{bmatrix}, \quad 
    X X\T = \begin{bmatrix} A A\T & AB \\ B\T A\T & B\T B \end{bmatrix}.
\end{equation*}
Computing $XX\T$ requires $M_{\stmm{gt}}(n)$ operations, but simultaneously computes three structured products and one general product $AB$. Thus $M_{\stmm{gt}}(n) = \Theta(M_{\stmm{gg}}(n)) = O(n^\omega)$ as well. The same argument extends to the other structured formats: asymptotically, $M_{\stmm{ab}}(n) = \Theta(M_{\stmm{gg}}(n)) = O(n^\omega)$ for all structures we consider.

Consequently, we can improve only the \emph{multiplicative constant}. We define the asymptotic complexity ratio
\begin{equation}
    \gamma_{\stmm{ab}} := \lim_{n \to \infty} \frac{M_{\stmm{ab}}(n)}{M_{\stmm{gg}}(n)},
    \label{eq:gamma}
\end{equation}
using $M_{\stmm{gg}}(n) = n^\omega$ as the base case. Our objective is to minimize $\gamma_{\stmm{ab}}$ for various structured formats by discovering low-rank tensor decompositions that enable efficient recursive algorithms.

\usubsec{Scheme Notation and Recursive Calls}
We extend the notation from Section~\ref{sec:algorithms} to distinguish different types of recursive calls within a scheme. A structured scheme is denoted
\begin{equation}
    \langle n_1, n_2, n_3 \colon r \rangle_{\stmm{ab}}^{(q_{\stmm{ab}}, q_{\stmm{ag}}, q_{\stmm{gb}})},
    \label{eq:notation}
\end{equation}
where $r$ is the total rank, and $(q_{\stmm{ab}}, q_{\stmm{ag}}, q_{\stmm{gb}})$ count recursive calls preserving both structures, left-only structure, and right-only structure, respectively. In this paper we focus on the square case $n_1 = n_2 = n_3$. For example, the scheme in Fig.~\ref{fig:gt-333} is written as $\langle 3,3,3 : 17 \rangle_{\stmm{gt}}^{(9,0,0)}$, it uses 17 scalar multiplications, of which 9 are recursive calls $A_j A_j\T$ and the remaining 8 are general multiplications.

\begin{table*}
    \caption{ 
    \justifying \textbf{Relative asymptotic complexity $M_{\stmm{ab}}/M_{\stmm{gg}}$}. Here $\stmm{a},\stmm{b}\in\{\stmm{g},\stmm{u},\stmm{l},\stmm{s},\stmm{k}\}$ denote \emph{general}, \emph{upper-triangular}, \emph{lower-triangular}, \emph{symmetric}, and \emph{skew-symmetric} matrices, respectively. The tag $\stmm{t}$ marks products with a transpose (e.g., $\stmm{gt}$ means $A\in\stmm{g}$ and we compute $AA\T$), for which only the upper-triangular part of the result is evaluated. ``This work'' reports the best ratios achieved by our schemes over integer coefficients ($\mathbb{Z}$) and rational coefficients ($\mathbb{Q}$). ``Baseline'' refers to algorithms that exploit structural zeros and matrix–vector product constructions for $\omega=3$, and their recursive application for $\omega=\log_2 7$. All ratios are with respect to $M_{\stmm{gg}}(n)=n^{\omega}$. Entries in bold indicate cases with the best known complexity. A dash indicates no improvement over $\Z$.
    }
    \begin{tabular}{l|c|llllllllllllll}
        \toprule
        & $\omega$ & \phantom{42}$\stmm{ug}$ & \phantom{42}$\stmm{sg}$ & \phantom{42}$\stmm{kg}$ & \phantom{42}$\stmm{gt}$ & \phantom{42}$\stmm{ut}$ & \phantom{42}$\stmm{st}$ & \phantom{42}$\stmm{kt}$ & \phantom{42}$\stmm{uu}$ & \phantom{42}$\stmm{us}$ & \phantom{42}$\stmm{uk}$ & \phantom{42}$\stmm{sk}$ & \phantom{42}$\stmm{ul}$  & \phantom{42}$\stmm{ss}$ & \phantom{42}$\stmm{kk}$
        \\ \midrule
        This work $(\Z)$ &  \multirow{3}{*}{$\log_2 7$} & 
        \tabiemph{0.595} & % ug        % n=4,lo,mod2,z,rank=34
        0.816 & % sg  % n=4,lo,mod2,z,rank=40
        0.816 & % kg        % n=4,lo,wg,mod2,z,rank=36
        \tabiemph{0.595} & % gt         % n=4,(10,0,0),mod2,z,rank=34
        \tabiemph{0.234} & % ut         % n=4,lo,(3,6,3),mod2,z,rank=19
        0.403 & % st
        0.416 & % kt
        \tabiemph{0.243} & % uu         % n=4,lo,(4,10,0),mod2,z,rank=
        \tabiemph{0.513} & % us
        \tabiemph{0.516} & % uk
        0.687 & % sk
        \tabiemph{0.425} & % ul
        0.653 & % ss              % n=4,lo,(0,1,0),mod3,q,rank=30
        0.687   % kk
        \\
        This work $(\Q)$ &  & 
        \hspace{4mm}- & % ug 
        \hspace{4mm}- & % sg 
        \tabiemph{0.806} & % kg 
        \hspace{4mm}- & % gt 
        \hspace{4mm}- & % ut 
        \tabiemph{0.360} & % st 
        \tabiemph{0.399} & % kt 
        \hspace{4mm}- & % uu 
        \hspace{4mm}- & % us 0.512
        \hspace{4mm}- & % uk 
        \tabiemph{0.615} & % sk 
        \hspace{4mm}- & % ul 
        \tabiemph{0.608} & % ss 
        \tabiemph{0.637} % kk 
        \\
        Baseline & & 
        0.615$^{\text{\cite{rybin_2025_10}}}$ & % ug
        0.816 & % sg
        0.816 & % kg
        % 0.634$^{\text{\cite{rybin_2025_05}}}$ & % gt
        0.615$^{\text{\cite{rybin_2025_10}}}$ & % gt
        0.306 & % ut
        0.588 & % st
        0.588 & % kt
        0.306 & % uu
        0.544 & % us (80/147, 4x4)
        0.544 & % uk
        0.799 & % sk
        0.467 & % ul (7/15, 2x2 base)
        0.799 & % ss
        0.799   % kk
        \\ \midrule
        Baseline & 3 & 
        0.444$^{\text{\cite{rybin_2025_10}}}$ & % ug
        0.500$^{\text{\cite{ye_2016_06}}}$ & % sg
        0.500$^{\text{\cite{fawzi_2022_10}}}$ & % kg
        0.444$^{\text{\cite{rybin_2025_10}}}$ & % gt
        0.167 & % ut
        0.333 & % st
        0.500 & % kt
        0.167 & % uu
          0.333 & % us
        0.333 & % uk
        0.500 & % sk
        0.333 & % ul
        0.500 & % ss
        0.500  % kk
        \\
        \bottomrule
    \end{tabular}
    \label{tab:relcomplex}
\end{table*}

\usubsec{Counting Structured Formats}
With input structures $\stmm{a},\stmm{b} \in \{\stmm{g}, \stmm{u}, \stmm{l}, \stmm{s}, \stmm{k}, \stmm{t}\}$, there are nominally $5 \times 6 = 30$ pairs $(\stmm{a}, \stmm{b})$ to consider.
However, many cases are equivalent due to symmetry. Since matrix transposition satisfies $C\T = B\T A\T$, we can relate schemes $\stmm{ab}$ and $\stmm{ba}$ by permuting tensor axes. Also $\stmm{lb}$ is equivalent to $\stmm{ub}$ via row and column permutation, so most cases involving $\stmm{l}$ can thus be reduced to corresponding $\stmm{u}$ cases. An exception is $\stmm{ul}$. The product of upper-triangular and lower-triangular matrices yields a \emph{general} matrix: $\stmm{ul} \to \stmm{g}$. In contrast, $\stmm{uu} \to \stmm{u}$ preserves upper-triangular structure. Consequently, $\stmm{ul}$ must be handled separately.

After accounting for symmetries, we obtain 15 non-trivial cases using the base structures $\{\stmm{g}, \stmm{u}, \stmm{l}, \stmm{s}, \stmm{k}, \stmm{t}\}$. Including the 5 additional variants where $\stmm{k}$ is replaced by $\stmm{w}$ for proper recursive treatment, we have 20 structured formats in total. Table~\ref{tab:relcomplex} lists asymptotic complexity ratios $\gamma_{\stmm{ab}}$ (with $\stmm{k}$-values computed via $\stmm{w}$-based schemes), Table~\ref{tab:schemes} reports base recursive schemes, and Table~\ref{tab:ranks} catalogs all discovered tensor ranks.

\usubsec{General Asymptotic Formula}
Returning to the example in Fig.~\ref{fig:gt-333}, the scheme is $\langle 3,3,3 : 17 \rangle_{\stmm{gt}}^{(9,0,0)}$, comprising 9 recursive calls of the form $A_j A_j\T$ ($\stmm{gt}$) and 8 general multiplications ($\stmm{gg}$). When applied recursively to larger matrices by partitioning into $3 \times 3$ blocks, this yields the recurrence
\begin{equation*}
    M_{\stmm{gt}}(n) = 9 M_{\stmm{gt}}(n/3) + 8  M_{\stmm{gg}}(n/3).
\end{equation*}
Using Strassen's algorithm as the base case ($\omega = \log_2 7$), this recurrence leads to $\gamma_{\stmm{gt}} \approx 0.623$ for the $3 \times 3$ scheme. 

For any structured scheme 
\eqref{eq:notation}
applied recursively with partition into $k \times k$ blocks, the asymptotic complexity takes the form $M_{\stmm{ab}}(n) = \gamma_{\stmm{ab}}\, n^\omega + o(n^\omega)$,
where the multiplicative factor $\gamma_{\stmm{ab}}$ is given by
\begin{equation}
    \gamma_{\stmm{ab}} = \frac{r - q_{\stmm{ab}} - q_{\stmm{ag}}(1 - \gamma_{\stmm{ag}}) - q_{\stmm{gb}}(1 - \gamma_{\stmm{gb}})}{k^\omega - q_{\stmm{ab}}}.
    \label{eq:factor}
\end{equation}
This formula assumes known auxiliary factors $\gamma_{\stmm{ag}}$, $\gamma_{\stmm{gb}}$ from their respective optimal schemes. The derivation is provided in the Appendix. As mentioned above, for Fig.~\ref{fig:gt-333} with $k=3$, $r=17$, $q_{\stmm{gt}}=9$, we have
$\gamma_{\stmm{gt}} = \frac{8}{3^{\omega}-9} \approx 0.623.$

\section{Flip Graph} \label{sec:flipgraph}

We can transform one correct tensor decomposition into another through local operations called \emph{flips}. A flip can be performed for any two terms sharing a common factor (Fig.~\ref{fig:flip-graph}a, top). When two terms share two factors, one of the resulting terms may become zero---an event called a \emph{reduction} (Fig.~\ref{fig:flip-graph}a, middle). These transformations generate a graph structure (Fig.~\ref{fig:flip-graph}b), first introduced in~\cite{kauers_2023_07}. It was shown in~\cite{kauers_2023_07} that the flip graph becomes connected once one allows the \emph{plus-transitions} introduced in~\cite{arai_2024_03}, namely compositions of an inverse reduction followed by a flip (Fig.~\ref{fig:flip-graph}a, bottom). Thus, with flips, reductions, and plus-transitions, one can transform any scheme into any other correct scheme.

Previous flip graph searches~\cite{kauers_2023_07,moosbauer_2025_02} were carried out over $\F_2$. Recent work~\cite{kauers_2025_10} also experimented with $\F_3$ and $\F_5$ in the meta flip graph and reported only matches of the best known rank bounds, without providing a public implementation of their search procedure. In this work we perform systematic flip graph searches over $\F_3$ with an open source implementation and, for the first time, obtain nontrivial schemes that intrinsically require the inverse of $2$, lift to $\Q$, and strictly improve on our $\F_2$ results.

\begin{table*}
    \caption{\justifying
        \textbf{Best recursive schemes for structured matrix multiplication.} 
        Each row shows: multiplicative factor $\gamma$, base size $n$, rank $r$, and recursive call distribution $(q_{\stmm{ab}}, q_{\stmm{ag}}, q_{\stmm{gb}})$. Upper section: schemes over $\Z$; lower section: schemes over $\Q$. Empty cells in the $\Q$ section indicate no improvement over $\Z$-schemes.
    }
    % $(\Z)$
    % $(\Q)$
    \begin{tabular}{c|ccccccccccccccc}
        \cmidrule[0.8pt]{1-15}
        & $\stmm{ug}$ & $\stmm{sg}$ & $\stmm{wg}$ & $\stmm{gt}$ & $\stmm{ut}$ & $\stmm{st}$ & $\stmm{wt}$ & $\stmm{uu}$ & $\stmm{us}$ & $\stmm{uw}$ & $\stmm{sw}$ & $\stmm{ul}$  & $\stmm{ss}$ & $\stmm{ww}$ & 
        \\ \cmidrule[0.5pt]{1-15}
        $\gamma$ & 
        {0.595} & % ug        % n=4,lo,mod2,z,rank=34
        0.816 & % sg  % n=4,lo,mod2,z,rank=40
        0.816 & % kg        % n=4,lo,wg,mod2,z,rank=36
        {0.595} & % gt         % n=4,(12,0,0),mod2,z,rank=34
        {0.234} & % ut         % n=4,lo,(4,5,6),mod2,z,rank=19
        0.403 & % st
        0.416 & % wt
        {0.243} & % uu         % n=4,lo,(4,10,0),mod2,z,rank=
        {0.513} & % us
        {0.516} & % uk
        0.687 & % sk
        {0.425} & % ul
        0.653 & % ss              % n=4,lo,(0,1,0),mod3,q,rank=30
        0.687 & % kk
        \multirow{6}{*}{$\left.\begin{array}{l} \\ \\ \\ \\ \\ \\ \end{array}\right\rbrace\Z$}
        \\
        $n$ &
        4 & % ug
        4 & % sg
        4 & % wg
        4 & % gt
        4 & % ut
        5 & % st
        4 & % wt
        4 & % uu
        5 & % us
        4 & % uw
        3 & % sw
        5 & % ul
        4 & % ss
        3 & % ww
        \\
        $r$ &
        34 & % ug
        40 & % sg
        40 & % wg
        34 & % gt
        19 & % ut
        39 & % st
        22 & % wt
        19 & % uu
        52 & % us
        29 & % uw
        15 & % sw
        47 & % ul
        32 & % ss
        15 & % ww
        \\
        $q_\stmm{ab}$ &
        12 & % ug
        & % sg
        & % wg
        12 & % gt
        4 & % ut
          & % st
          & % wt
        4 & % uu
        1 & % us
        1 & % uw
        & % sw
        2 & % ul
        & % ss
        & % ww
        \\
        $q_\stmm{ag}$ &
        & % ug
        & % sg
        & % wg
        & % gt
        5 & % ut
         & % st
        & % wt
        10 & % uu
        11 & % us
        8 & % uw
        & % sw
        17& % ul
        & % ss
        & % ww
        \\
        $q_\stmm{gb}$ &
        & % ug
        & % sg
        & % wg
        & % gt
        6 & % ut
        5 & % st
        4 & % wt
        & % uu
        & % us
        & % uw
        & % sw
        & % ul
        & % ss
        & % ww
        \\ \cmidrule[0.5pt]{1-15}
        $\gamma$ &  
         & % ug 
         & % sg 
        {0.806} & % kg 
         & % gt 
         & % ut 
        {0.360} & % st 
        {0.399} & % wt 
         & % uu 
        0.512 & % us 
         & % uk 
        {0.615} & % sk 
         & % ul 
        {0.608} & % ss 
        {0.637} & % kk
        \multirow{6}{*}{$\left.\begin{array}{l} \\ \\ \\ \\ \\ \\ \end{array}\right\rbrace\Q$}
        \\
        $n$ &
        & % ug
        & % sg
        3 & % wg
        & % gt
        & % ut
        5 & % st
        4 & % wt
        & % uu
        4 & % us
        & % uw
        3 & % sw
        & % ul
        4 & % ss
        3& % ww
        \\
        $r$ &
        & % ug
        & % sg
        18 & % wg
        & % gt
        & % ut
        35 & % st
        22 & % wt
        & % uu
        28 & % us
        & % uw
        14 & % sw
        & % ul
        30 & % ss
        15 & % ww
        \\
        $q_\stmm{ab}$ &
        & % ug
        & % sg
        2 & % wg
        & % gt
        & % ut
        & % st
        & % wt
        & % uu
        1 & % us
        & % uw
        & % sw
        & % ul
        & % ss
        3 & % ww
        \\
        $q_\stmm{ag}$ &
        & % ug
        & % sg
        & % wg
        & % gt
        & % ut
        & % st
         & % wt
        & % uu
        6 & % us
        & % uw
        2 & % sw
        & % ul
        1 & % ss
        & % ww
        \\
        $q_\stmm{gb}$ &
        & % ug
        & % sg
        & % wg
        & % gt
        & % ut
        5 & % st
        6 & % wt
        & % uu
        & % us
        & % uw
        1 & % sw
        & % ul
        & % ss
        & % ww
        \\
        \cmidrule[0.8pt]{1-15}
    \end{tabular}
    \label{tab:schemes}
\end{table*}

Regarding the search procedure itself, the general definition of reduction~\cite{kauers_2023_07,arai_2024_03} requires checking for arbitrary linear dependencies across terms with a shared factor. However, such general dependencies rarely occur in practice~\cite{moosbauer_2025_02}, and explicit reduction checks are computationally expensive, approximately 5 times slower than performing flips alone in our experiments. We therefore adopt the simplified approach from~\cite{moosbauer_2025_02}, performing only flips and plus-transitions, without explicit reduction checks. Reductions are discovered implicitly when they occur during random flips.

% Search algorithm
Our search procedure follows the framework from~\cite{kauers_2023_07}, but simplified following~\cite{moosbauer_2025_02}. We maintain a pool of schemes at the current rank $r$. At each iteration, we select a random scheme from the pool and perform a random walk, attempting to reach rank $r-1$ or lower. Each step of the walk consists of selecting a random flip from the current scheme. If the walk stagnates, making no progress for $P$ consecutive steps, we perform a plus-transition to escape the potential local plateau. We terminate the walk after at most $L$ steps, or earlier if we successfully reach a lower rank. Once we accumulate a pool of $S$ schemes at rank $r-1$, we repeat the process targeting rank $r-2$.

For all experiments reported in this paper, we use the parameters specified above: walk length limit $L = 10^6$, stagnation threshold $P = 5 \times 10^4$, and pool size $S=10^4$. This randomized procedure proves remarkably effective at discovering low-rank schemes, as demonstrated by our results in Table~\ref{tab:relcomplex}.

% Challenge: local traps and importance of pool diversity
Reductions can lead to regions where further descent requires plus-transitions, as illustrated in Fig.~\ref{fig:flip-graph}b. A scheme at rank $r$ may reduce to rank $r-1$, but then become isolated in a connected component with no further reductions. This is why maintaining a diverse pool of schemes is essential: different schemes may lead to different regions of the graph at rank $r-1$, some of which may have paths to lower ranks. The plus-transition operation provides an additional mechanism to escape such local minima by temporarily increasing complexity before finding alternative descent paths. Without both the pool diversity and plus-transitions, random walks would frequently terminate at suboptimal ranks.

% Hensel lifting 
After obtaining a pool of $10^4$ schemes over $\F_2$ or $\F_3$, we apply Hensel lifting \cite{kauers_2023_07} to identify which schemes can be lifted to $\Z$ or $\Q$. Starting from a scheme valid modulo $p \in \{2,3\}$, Hensel lifting constructs a $p$-adic approximation by iteratively refining the solution modulo $p^k$ for increasing $k$. At each step, the refinement requires solving a linear system over $\Z_p$. If the scheme is $\Z_p$-specific and cannot be lifted, this system has no solution. In this work, we perform $k=10$ lifting steps to obtain a sufficiently accurate $p$-adic approximation. We then apply rational reconstruction to recover a candidate scheme with coefficients in $\Z$ or $\Q$. 

Table~\ref{tab:ranks} summarizes the best ranks achieved through this search procedure for all structured matrix multiplication formats $\stmm{ab}$ with $n \in \{2,3,4,5\}$, reporting results separately for $\F_2$, $\F_3$, $\Z$, and $\Q$. The schemes over $\Z$ and $\Q$ are obtained through Hensel lifting from the corresponding finite field schemes.

\section{Identifying Recursive Schemes} \label{sec:recursive}

Among the schemes successfully lifted to $\Z$ or $\Q$, we identify those that enable recursive block matrix multiplication. For a scheme of format $\stmm{ab}$, we interpret each rank-1 term $U \otimes V \otimes W$ as operating on matrix blocks. A term contributes a \emph{recursive call} if its input matrices inherit the structure: specifically, if $U$ and $V$ are nonzero only on diagonal blocks, the corresponding multiplication can be computed recursively. Depending on which factors preserve structure, we classify recursive calls into three types: $q_{\stmm{ab}}$ (both inputs structured), $q_{\stmm{ag}}$ (left-structured, right general), and $q_{\stmm{gb}}$ (left general, right-structured).

For transpose product schemes $\stmm{t}$, two criteria for identifying recursive calls arise naturally. The first considers a term recursive if $U = V$ after lifting: since the left and right inputs are identical, the result is symmetric, so only the upper-triangular part needs to be computed. The second considers a term recursive if $W$ contributes only to diagonal output blocks, which are $A_i A_i\T$ subproblems. Both criteria are valid and yield different recursive call counts for the same scheme. We evaluate schemes under both criteria and select the one yielding better asymptotic complexity for each $(n, \stmm{at})$ pair. In practice, neither criterion uniformly dominates: for instance, the $3 \times 3$ scheme in Fig.~\ref{fig:gt-333} achieves 9 recursive calls only under the first criterion, while the $4 \times 4$ scheme in Fig.~\ref{fig:gt-444} achieves 12 recursive calls under the second.

Each scheme is thus characterized by a triple $(q_{\stmm{ab}}, q_{\stmm{ag}}, q_{\stmm{gb}})$ counting its recursive calls of each type. We identify the Pareto frontier: schemes that are not dominated by any other scheme having all three counts greater or equal with at least one strictly greater. Among Pareto-optimal schemes, we prioritize those over $\Z$, and only include schemes over $\Q$ if they dominate all integer schemes. Within each coefficient domain, we select the scheme with the smallest denominator (for rational coefficients) and, as a tiebreaker, the fewest nonzero entries in the decomposition $(U, V, W)$. Table~\ref{tab:relcomplex} reports the best factors $\gamma$ achieved for each structured format, and Table~\ref{tab:schemes} provides the corresponding scheme parameters.

For transpose product formats, we employ an additional technique inspired by the RXTX algorithm~\cite{rybin_2025_05}. Preliminary experiments for $n \in \{2,3,4,5\}$ revealed that the corner elements of the result matrix (corresponding to the upper-left and lower-right diagonal blocks) can often be computed recursively. We therefore perform a specialized search starting from a tensor with removed components corresponding to these corner elements, forcing $2n-2$ terms of the decomposition to compute off-diagonal outputs and the remaining $2$ to handle the corners via $\stmm{at}$ recursion. Fig.~\ref{fig:gt-333} illustrates this structure for the $3 \times 3$ case: beyond the corner blocks, other diagonal elements also benefit from recursively computed intermediate terms, though forcing all outputs to be fully recursive would increase the rank.

\section{Results} \label{sec:results}

\begingroup

\setlength{\tabcolsep}{5pt}

\begin{table*}
    \caption{\justifying
        \textbf{Ranks of discovered schemes for structured matrix multiplication.} 
        For each format and block size $n$, we report the best found ranks over $\F_2$, $\F_3$, $\Z$, and $\Q$. The $\text{nnz}$ column shows the naive tensor rank (number of nonzeros in $T_{ijk}$). Schemes over $\Z$ and $\Q$ are obtained by Hensel lifting from finite field schemes. A dash indicates no improvement over previous columns. 
        Bold entries highlight cases where $\F_2$ scheme could not be lifted.
    }
    \begin{tabular}{c|ccccc|ccccc|ccccc|ccccc}
        \toprule
        &\multicolumn{5}{c|}{$n=2$}&\multicolumn{5}{c|}{$n=3$}&\multicolumn{5}{c|}{$n=4$}&\multicolumn{5}{c}{$n=5$}\\
        & nnz & $\F_2$ & $\F_3$ & $\Z$ & $\Q$ & nnz & $\F_2$ & $\F_3$ & $\Z$ & $\Q$ & nnz & $\F_2$ & $\F_3$ & $\Z$ & $\Q$ & nnz & $\F_2$ & $\F_3$ & $\Z$ & $\Q$ \\
        \midrule
        %             nnz  Z2   Z3   Z    Q  | nnz  Z2   Z3   Z    Q  | nnz  Z2   Z3   Z    Q  | nnz  Z2   Z3   Z    Q    
        $\stmm{gg}$ & 8  & 7  & 7  & 7  & -  & 27 & 23 & 23 & 23 & -  & 64 & \textbf{47} & 49 & 49 & -  & 125& 97 & 124& 97 & -  \\
        $\stmm{ug}$ & 6  & -  & -  & -  & -  & 18 & 17 & 17 & 17 & -  & 40 & 34 & 34 & 34 & -  & 75 & 63 & 70 & 63 & -  \\
        $\stmm{sg}$ & 8  & 6  & 6  & 6  & -  & 27 & 18 & 18 & 18 & -  & 64 & 40 & 40 & 40 & -  & 125& 75 & 75 & 75 & -  \\
        $\stmm{kg}$ & 4  & -  & -  & -  & -  & 18 & 15 & 14 & 15 & 14 & 48 & 36 & 36 & 36 & -  & 100& 70 & 75 & 70 & -  \\
        $\stmm{gt}$ & 6  & -  & -  & -  & -  & 18 & 17 & 17 & 17 & -  & 40 & 34 & 34 & 34 & -  & 75 & 63 & 74 & 63 & -  \\
        $\stmm{ut}$ & 4  & -  & -  & -  & -  & 10 & -  & -  & -  & -  & 20 & 19 & 19 & 19 & -  & 35 & 32 & 32 & 32 & -  \\
        $\stmm{st}$ & 6  & 4  & 4  & 4  & -  & 18 & 11 & 10 & 11 & 10 & 40 & 22 & 20 & 22 & 20 & 75 & 39 & 35 & 39 & 35 \\
        $\stmm{kt}$ & 2  & 1  & 1  & 1  & -  & 9  & 6  & 6  & 6  & -  & 24 & 15 & 15 & 15 & 15 & 50 & \textbf{29} & 30 & 30 & -  \\
        $\stmm{uu}$ & 4  & -  & -  & -  & -  & 10 & -  & -  & -  & -  & 20 & 19 & 19 & 19 & 19 & 35 & 32 & 32 & 32 & -  \\
        $\stmm{us}$ & 6  & 5  & 5  & 5  & -  & 18 & 14 & 14 & 14 & -  & 40 & 29 & 28 & 29 & 28 & 75 & 52 & 54 & 52 & -  \\
        $\stmm{uk}$ & 3  & -  & -  & -  & -  & 12 & 11 & 10 & 11 & 10 & 30 & 24 & 24 & 24 & 24 & 60 & 45 & 49 & 45 & -  \\
        $\stmm{sk}$ & 4  & 3  & 3  & 3  & -  & 18 & 11 & 11 & 11 & -  & 48 & 26 & 24 & 26 & 24 & 100& 50 & 57 & 50 & -  \\
        $\stmm{ul}$ & 5  & -  & -  & -  & -  & 14 & 13 & 13 & 13 & -  & 30 & 27 & 27 & 27 & -  & 55 & 47 & 50 & 47 & -  \\
        $\stmm{ss}$ & 8  & 6  & 5  & 6  & 5  & 27 & 15 & 14 & 15 & 14 & 64 & 32 & 30 & 32 & 30 & 125& 59 & 62 & 59 & -  \\
        $\stmm{kk}$ & 2  & 1  & 1  & 1  & -  & 12 & \textbf{8}  & 9  & 9  & 9  & 36 & 21 & 20 & 22 & 20 & 80 & \textbf{45} & 50 & 48 & -  \\
        $\stmm{wg}$ & 8  & 6  & 6  & 6  & -  & 27 & 18 & 18 & 18 & -  & 64 & 40 & 40 & 40 & -  & 125& 75 & 75 & 75 & -  \\
        $\stmm{wt}$ & 6  & 4  & 4  & 4  & -  & 18 & 11 & 11 & 11 & -  & 40 & 22 & 22 & 22 & -  & 75 & \textbf{39} & 40 & 44 & 40 \\
        $\stmm{uw}$ & 6  & 5  & 5  & 5  & -  & 18 & 14 & 14 & 14 & -  & 40 & 29 & 29 & 29 & -  & 75 & \textbf{52} & 54 & 54 & -  \\
        $\stmm{sw}$ & 8  & 6  & 5  & 6  & 5  & 27 & 15 & 14 & 15 & 14 & 64 & 32 & 31 & 35 & 31 & 125& \textbf{59} & 64 & 64 & -  \\
        $\stmm{ww}$ & 8  & 6  & 5  & 6  & 5  & 27 & 15 & 15 & 15 & -  & 64 & \textbf{32} & 33 & 35 & 33 & 125& \textbf{59} & 68 & 68 & -  \\
        \bottomrule
    \end{tabular}
    \label{tab:ranks}
\end{table*}

\endgroup

We present a systematic exploration of structured matrix multiplication across all $15$ (and auxiliary $5$ with $\stmm{w}$) distinct format combinations arising from input structures $\stmm{a} \in \{\stmm{g}, \stmm{u}, \stmm{l}, \stmm{s}, \stmm{k}\}$ and $\stmm{b} \in \{\stmm{g}, \stmm{u}, \stmm{l}, \stmm{s}, \stmm{k}, \stmm{t}\}$ after accounting for symmetries. Our flip graph search over $\F_2$ and $\F_3$ yielded improved asymptotic complexity factors for 13 of these 15 cases, with only $\stmm{gg}$ and $\stmm{sg}$ remaining. Table~\ref{tab:relcomplex} reports the achieved ratios $\gamma_{\stmm{ab}}$ defined in~\eqref{eq:gamma}, comparing against the best previously known algorithms. For base sizes $n \in \{2,3,4,5\}$, we provide a complete catalog of discovered tensor ranks across coefficient domains $\F_2$, $\F_3$, $\Z$, and $\Q$, covering 80 distinct tensors in total (Table~\ref{tab:ranks}). These structured products correspond primarily to Level~3 BLAS operations, notably symmetric rank-$k$ updates (SYRK -- our  $\stmm{gt}$) and triangular matrix multiplication (TRMM -- our $\stmm{ug}$).

A notable example is a SYRK scheme for computing $AA\T$, denoted by  $\langle 4,4,4 : 34 \rangle_{\stmm{gt}}^{(12,0,0)}$, which attains $\gamma_{\stmm{gt}} = 22/37 \approx 0.595$, improving the multiplicative factor from the previous best of $8/13 \approx 0.615$ \cite{rybin_2025_05,rybin_2025_10}. The complete decomposition is provided explicitly in Fig.~\ref{fig:gt-444}. 

Our extension to flip graph search over $\F_3$ enabled discovery of schemes fundamentally requiring the inverse of 2. While we did not independently recover the rank-48 scheme for $\langle 4,4,4 \rangle_{\stmm{gg}}$ reported in recent work~\cite{kaporin_2024_09,novikov_2025_06,dumas_2025_07}, our $\F_3$ search discovered several other schemes requiring the inverse of 2. We found $\langle 2,2,2 : 5 \rangle_{\stmm{ss}}$ for symmetric-symmetric multiplication and $\langle 3,3,3 : 14 \rangle_{\stmm{kg}}$ for skew-symmetric times general, improving upon previous best ranks of 6~\cite{ye_2018_02} and 15~\cite{fawzi_2022_10} obtained via matrix-vector constructions. Both schemes lift successfully from $\F_3$ to $\Q$ via Hensel lifting. 

Our computational setup used 48-core Intel Xeon Gold 6246 nodes; each $(n,\stmm{ab},\F_{\{2,3\}})$ combination was searched on a single node with a 24-hour time limit. The complete search required 1007 core-days. As concrete examples of search efficiency, the $\langle 4, 4, 4 : 34 \rangle_{\text{gt}}^{(12,0,0)}$ scheme (Fig.~4) was found in 10 minutes of wall-clock time.
% , and the $\langle 4, 4, 4 : 34 \rangle_{\text{ug}}^{(12,0,0)}$ scheme (Fig.~6) in 14 minutes

We also encountered several schemes that do not lift beyond $\F_2$, analogous to the known $\langle 4,4,4:47 \rangle_{\stmm{gt}}$ case. Examples include $\langle 5,5,5:45 \rangle_{\stmm{kk}}$ and $\langle 5,5,5:52 \rangle_{\stmm{uw}}$, among others highlighted in Table~\ref{tab:ranks}. Notably, among $\F_3$ schemes it was always possible to find a liftable scheme.

For the baseline comparisons in Table~\ref{tab:relcomplex}, the $\omega=3$ entries represent classical constructions exploiting structural zeros and matrix-vector products from~\cite{ye_2018_02,ye_2016_06,fawzi_2022_10}. The $\omega=\log_2 7$ entries correspond to recursive schemes built atop these constructions, adapted to Strassen's exponent. Where no explicit reference is provided, the listed ``baseline'' factor indicates what would be achievable without this work, constructed via standard techniques. The $\stmm{ug}$ example derivation is provided in the Appendix.

As evident from equation~\eqref{eq:factor}, all reported asymptotic factors $\gamma_{\stmm{ab}}$ depend on the choice of matrix multiplication exponent $\omega$. We focus on $\omega = \log_2 7$ throughout for ease of comparison with prior work, but our schemes remain applicable for any value of $\omega$; different exponents will alter the numerical values of $\gamma_{\stmm{ab}}$ and potentially change which base schemes yield optimal recursive algorithms. Exact values of $\gamma_{\stmm{ab}}$ for all discovered schemes can be recovered by substituting parameters from Table~\ref{tab:schemes} into~\eqref{eq:factor}.

Among the 15 structured formats, two cases ($\stmm{gg}$ and $\stmm{sg}$) remain at their previously known bounds. For general matrix multiplication $\stmm{gg}$, this is expected: it has been the primary focus of extensive research. Improving asymptotic bounds for $\stmm{gg}$ was not a goal of this work. For symmetric-general multiplication ($\stmm{sg}$), the lack of improvement  likely reflects insufficient exploration rather than a fundamental limitation—larger base sizes or more extensive search would presumably yield better schemes.

All $\Z$ and $\Q$ schemes reported in Tables~\ref{tab:schemes} and \ref{tab:ranks} are released together with the full search code and runnable examples in a public repository: \href{http://github.com/khoruzhii/flip-cpd}{\texttt{github.com/khoruzhii/flip-cpd}}. The repository provides a C++ implementation of our flip graph search pipeline for CP decomposition of arbitrary 3-way tensors over $\F_2$ and $\F_3$, with integrated Hensel lifting and rational reconstruction. The implementation sustains approximately $5 \times 10^6$ flips per second per thread on standard hardware (see Appendix for details). The repository also includes utilities to reproduce the results, verification scripts, and examples showing how to run the search for a new tensor.

\section{Discussion} \label{sec:discussion}
The flip graph search methodology demonstrated remarkable efficiency: we discovered improved bounds for most of structured matrix multiplication formats and  obtained thousands of schemes across different tensor sizes. This success suggests several promising directions for future research.

The flip graph framework naturally extends beyond matrix multiplication to other bilinear mappings. Polynomial multiplication, multiplication in algebras of complex numbers, quaternions \cite{dumas_2021_01}, and octonions can all be formulated as bilinear schemes where the same search methodology applies without algorithmic modifications, if one wants to minimize number of real multiplications. 

Beyond direct applications to new problem domains, several recent modifications of the flip graph method enable the exploration of related complexity measures; for example, the commutative variant \cite{wood_2025_06} and the approximate-scheme approach \cite{moosbauer_2023_09}. The latter provides upper bounds on border rank. Another direction involves working over different base fields: while we searched over $\mathbb{F}_2$ and $\mathbb{F}_3$, fields such as $\mathbb{F}_3[i]$ could enable discovery of schemes minimizing complex multiplications.

Our approach optimizes for minimal rank, then filters discovered schemes by their recursive call distributions. However, practical algorithm design requires simultaneous optimization of multiple objectives: rank, number of recursive calls, numerical stability and total addition count. We focus on the algebraic aspects of scheme discovery; the design of implementations is a problem that we view as complementary future work. We introduced the zero-corners technique as a partial solution, forcing certain recursive patterns by zeroing specific tensor components. More systematic approaches could involve fixing particular vectors in the decomposition factors $U$, $V$, $W$, or employing annealing-like methods to gradually adjust search priorities. However, the fundamental question remains open: how can we navigate the flip graph to simultaneously optimize these objectives? Developing multi-objective search strategies on flip graphs represents an important direction for discovering schemes with better practical performance characteristics.

Transitioning to flip graphs already dramatically reduced the search space by restricting attention to provably correct schemes. Random walk approach, while successful, performs undirected exploration within this space. The natural next step involves applying reinforcement learning \cite{fawzi_2022_10} or diffusion methods \cite{chervov_2025_02} to guide the search toward low-rank regions of the flip graph. Such directed search on the restricted space of correct schemes could combine the computational efficiency of flip graph methods with the adaptive exploration capabilities of machine learning, potentially enabling discovery of low-rank decompositions for larger tensor formats that remain intractable for purely random exploration.

\phantom{42}

\noindent
\textbf{Acknowledgments}. 
We would like to express our sincere gratitude to Lieven De Lathauwer and Charlotte Vermeylen from KU Leuven for the fruitful and engaging discussions. We also thank Dmitry Rybin for valuable comments and helpful suggestions. Their input and perspectives were greatly appreciated. This research was supported by the DFG Cluster of Excellence MATH+ (EXC-2046/1, project id 390685689) funded by the Deutsche Forschungsgemeinschaft (DFG), as well as by the National High-Performance Computing (NHR) network.

\bibliographystyle{apsrev4-2}
\bibliography{stmm.bib}

\begin{figure*}

\begin{minipage}[t]{0.37\textwidth}
    \raggedright
    General multiplications
    \small{
    \setlength{\jot}{2pt} 
    \begin{equation*}
        \begin{aligned}
            m_{1}  &= (A_3^2)\blue{(A_4^2 + A_4^3)\T} \\
            m_{2}  &= (A_1^4 + A_2^4)\blue{(A_3^4 - A_4^4)\T} \\
            m_{3}  &= (A_1^1 + A_2^3)\blue{(A_3^1 + A_4^3)\T} \\
            m_{4}  &= (A_1^3 + A_2^3)\blue{(A_3^3 - A_4^3)\T} \\
            m_{5}  &= (A_1^2 + A_2^2)\blue{(A_3^2 - A_4^2)\T} \\
            m_{6}  &= (A_1^4 + A_2^2)\blue{(A_3^2 + A_4^4)\T} \\
            m_{7}  &= (A_1^1 + A_2^1)\blue{(A_3^1 - A_4^1)\T} \\
            m_{8}  &= (A_2^2)\blue{(A_2^1 + A_2^2 - A_3^2 - A_4^4)\T} \\
            m_{9}  &= (A_1^4)\blue{(A_2^3 - A_2^4 - A_3^2 - A_4^4)\T} \\
            m_{10} &= (A_1^1 + A_2^1 - A_3^1)\blue{(A_4^1 + A_4^4)\T} \\
            m_{11} &= (A_1^1 + A_2^3 + A_3^2 - A_3^3)\blue{(A_4^3)\T} \\
            m_{12} &= (A_1^1 - A_1^2 + A_1^3 + A_1^4)\blue{(A_2^3)\T} \\
            m_{13} &= (A_1^1 + A_2^3 - A_3^1 + A_3^4)\blue{(A_3^1)\T} \\
            m_{14} &= (A_1^1 + A_2^2)\blue{(A_2^1 + A_2^2 - A_3^2 + A_4^2)\T} \\
            m_{15} &= (A_1^1 - A_1^2)\blue{(A_2^2 + A_2^3 - A_3^2 + A_4^2)\T} \\
            m_{16} &= (A_1^4 - A_2^3)\blue{(A_2^3 - A_2^4 + A_3^4 - A_4^4)\T} \\
            m_{17} &= (A_1^1 + A_2^1 - A_3^1 + A_3^4)\blue{(A_3^1 + A_4^4)\T} \\
            m_{18} &= (A_1^3 + A_2^3 + A_3^2 - A_3^3)\blue{(A_3^2 + A_4^3)\T} \\
            m_{19} &= (A_1^1)\blue{(A_2^1 + A_2^2 - A_3^2 + A_4^1 + A_4^2 + A_4^3)\T} \\
            m_{20} &= (A_2^3)\blue{(A_2^3 - A_2^4 - A_3^1 - A_3^3 + A_3^4 - A_4^4)\T} \\
            m_{21} &= (A_1^2 - A_1^3 - A_1^4 - A_2^3 - A_3^2 + A_3^3)\blue{(A_3^2)\T} \\
            m_{22} &= (A_1^1 + A_2^1 + A_2^2 - A_2^4 - A_3^1 + A_3^4)\blue{(A_4^4)\T} \\
        \end{aligned}
    \end{equation*}
    }
\end{minipage}
\hfill
\begin{minipage}[t]{0.62\textwidth}
    \raggedright
    Block layout
    \small{
    \begin{equation*}
        \begin{pmatrix}
            C_{1}^{1} & C_{1}^{2} & C_{1}^{3} & C_{1}^{4} \\
            \grey{C_{2}^{1}} & C_{2}^{2} & C_{2}^{3} & C_{2}^{4} \\
            \grey{C_{3}^{1}} & \grey{C_{3}^{2}} & C_{3}^{3} & C_{3}^{4} \\
            \grey{C_{4}^{1}} & \grey{C_{4}^{2}} & \grey{C_{4}^{3}} & C_{4}^{4}
        \end{pmatrix}
        =
        \begin{pmatrix}
            A_{1}^{1} & A_{1}^{2} & A_{1}^{3} & A_{1}^{4} \\
            A_{2}^{1} & A_{2}^{2} & A_{2}^{3} & A_{2}^{4} \\
            A_{3}^{1} & A_{3}^{2} & A_{3}^{3} & A_{3}^{4} \\
            A_{4}^{1} & A_{4}^{2} & A_{4}^{3} & A_{4}^{4}
        \end{pmatrix}
        \blue{
        \begin{pmatrix}
            A_{1}^{1} & A_{1}^{2} & A_{1}^{3} & A_{1}^{4} \\
            A_{2}^{1} & A_{2}^{2} & A_{2}^{3} & A_{2}^{4} \\
            A_{3}^{1} & A_{3}^{2} & A_{3}^{3} & A_{3}^{4} \\
            A_{4}^{1} & A_{4}^{2} & A_{4}^{3} & A_{4}^{4}
        \end{pmatrix}^{\hspace{-1.5mm}\textnormal{T}}
        }
    \end{equation*}
    }

    \phantom{42}

    Recursive multiplications
    \small{
    \begin{equation*}
        \begin{aligned}
            \reccall{r_{1}} &= (A_1^1)\blue{(A_1^1)\T} \\
            \reccall{r_{2}} &= (A_1^2)\blue{(A_1^2)\T} \\
            \reccall{r_{3}} &= (A_1^3)\blue{(A_1^3)\T} \\
            \reccall{r_{4}} &= (A_1^4)\blue{(A_1^4)\T} \\
        \end{aligned}
        \hspace{0.3cm} 
        \begin{aligned}
            \reccall{r_{5}} &= (A_4^1)\blue{(A_4^1)\T} \\
            \reccall{r_{6}} &= (A_4^2)\blue{(A_4^2)\T} \\
            \reccall{r_{7}} &= (A_4^3)\blue{(A_4^3)\T} \\
            \reccall{r_{8}} &= (A_4^4)\blue{(A_4^4)\T} \\
        \end{aligned}
        \hspace{0.3cm} 
        \begin{aligned}
            \reccall{r_{9}}  &= (A_2^3 + A_2^4)\blue{(A_2^1 + A_2^4 - A_3^4 + A_4^4)\T} \\
            \reccall{r_{10}} &= (A_2^1 - A_2^2 - A_2^3 - A_2^4)\blue{(A_2^1)\T} \\
            \reccall{r_{11}} &= (A_3^4)\blue{(A_3^1 + A_3^4)\T} \\
            \reccall{r_{12}} &= (A_1^3 + A_2^3 - A_3^3)\blue{(A_3^2 + A_3^3)\T} \\
        \end{aligned}
    \end{equation*}
    }

    \phantom{42}

    Output
    \small{
    \begin{equation*}
        \begin{aligned}
            C_1^1  &= \reccall{r_{1} + r_{2} + r_{3} + r_{4}} \\
            C_1^2  &= m_{5} + m_{14} + m_{12} - m_{8} - m_{6} - m_{15} - m_{9} \\
            C_1^3  &= -m_{11} + m_{3} + m_{20} + m_{4} + m_{18} - m_{9} + m_{16} + m_{21} \\
            C_1^4  &= -m_{11} - m_{5} + m_{19} - m_{14} + m_{8} + m_{6} + m_{18} + m_{21} \\
            C_2^2  &= \reccall{r_{9} + r_{10}} + m_{2} + m_{8} + m_{6} +  m_{9} - m_{16} \\
            C_2^3  &= m_{2} - m_{20} + m_{17} + m_{6} - m_{13} + m_{9} - m_{16} - m_{22} \\
            C_2^4  &= m_{3} - m_{19} + m_{14} + m_{17} - m_{8} - m_{13} - m_{7} - m_{22} \\
            C_3^3  &= \reccall{r_{11}-r_{12}} - m_{11} + m_{3} + m_{4} - m_{13}  + m_{18} \\
            C_3^4  &= -m_{11} + m_{3} - m_{10} + m_{1} + m_{17} - m_{13} - m_{7} \\
            C_4^4  &= \reccall{r_{5} + r_{6} + r_{7} + r_{8}} \\
        \end{aligned}
    \end{equation*}
    }
\end{minipage}

\vspace{-0.2cm}
\caption{
    \justifying
    $\langle 4, 4, 4 \colon 34 \rangle_\stmm{gt}^{(12,0,0)}$.
    Structured matrix multiplication \(C = A A^{\mathrm T}\) for a \(4\times4\) block with symmetric \(C\).
    Coefficients lie in \(\mathbb{Z}\). 
    Operation count: 34 multiplications ($12\,\stmm{gt} + 22\,\stmm{gg}$) and 141 additions.
}
\label{fig:gt-444}
\end{figure*}
\begin{figure*}

\begin{minipage}[t]{0.46\textwidth}
    \raggedright
    General multiplications
    \small{
    \begin{equation*}
        \begin{aligned}
            m_{1}  &= (A_{1}^{2} - A_{1}^{3} + A_{2}^{3})\blue{(B_{1}^{1} - B_{1}^{2} + B_{2}^{3} - B_{3}^{1} + B_{3}^{2})}\inverseoftwo{/2} \\
            m_{2}  &= (A_{1}^{2} - A_{1}^{3} + A_{2}^{3})\blue{(B_{1}^{1} + B_{2}^{3} + B_{3}^{2} - B_{3}^{3})}\inverseoftwo{/2} \\
            m_{3}  &= (A_{1}^{2} + A_{2}^{3})\blue{(B_{1}^{1} - B_{1}^{2} + B_{2}^{1} - B_{2}^{2} - B_{2}^{3} - B_{3}^{1} + B_{3}^{2})}\inverseoftwo{/2} \\
            m_{4}  &= (A_{1}^{2})\blue{(B_{1}^{1} + B_{1}^{2} - B_{1}^{3} + B_{2}^{2})} \\
            m_{5}  &= (A_{1}^{2} - A_{2}^{3})\blue{(B_{1}^{1} + B_{2}^{1} - B_{2}^{3} - B_{3}^{2} + B_{3}^{3})}\inverseoftwo{/2} \\
            m_{6}  &= (A_{1}^{2} + A_{1}^{3} - A_{2}^{3})\blue{(B_{1}^{1} + B_{2}^{3} - B_{3}^{2} + B_{3}^{3})}\inverseoftwo{/2} \\
            m_{7}  &= (A_{1}^{3} - A_{2}^{3})\blue{(B_{2}^{3} + B_{3}^{1} + B_{3}^{2} - B_{3}^{3})}\inverseoftwo{/2} \\
            m_{8}  &= (A_{1}^{2} + A_{1}^{3} - A_{2}^{3})\blue{(B_{1}^{2} - B_{3}^{1} + B_{3}^{3})}\inverseoftwo{/2} \\
            m_{9}  &= (A_{1}^{2} + A_{2}^{3})\blue{(B_{1}^{2} + B_{2}^{2} + B_{3}^{1} - B_{3}^{3})}\inverseoftwo{/2} \\
            m_{10} &= (A_{1}^{3})\blue{(2B_{1}^{3} + B_{2}^{3})}\inverseoftwo{/2} \\
            m_{11} &= (A_{1}^{2} - A_{2}^{3})\blue{(B_{1}^{1} - B_{1}^{2} + B_{2}^{1} - B_{2}^{2} - B_{2}^{3} + B_{3}^{1} - B_{3}^{2})}\inverseoftwo{/2} \\
            m_{12} &= (A_{1}^{3} - 2A_{2}^{3})\blue{(2B_{2}^{3} + B_{3}^{1} + B_{3}^{2} - B_{3}^{3})}\inverseoftwo{/2} \\
            m_{13} &= (A_{1}^{3})\blue{(2B_{1}^{3} + B_{2}^{3} - B_{3}^{1} + B_{3}^{2} + B_{3}^{3})}\inverseoftwo{/2} \\
            m_{14} &= (A_{1}^{2})\blue{(B_{1}^{1} + B_{1}^{2} - B_{1}^{3} - B_{2}^{1} + 2B_{2}^{2} + B_{2}^{3})}
        \end{aligned}
    \end{equation*}
    }
\end{minipage}
\hfill
\begin{minipage}[t]{0.52\textwidth}
    \raggedright
    Block layout
    \small{
    \begin{equation*}
        \begin{pmatrix}
            C_{1}^{1} & C_{1}^{2} & C_{1}^{3}  \\
            C_{2}^{1} & C_{2}^{2} & C_{2}^{3}  \\
            C_{3}^{1} & C_{3}^{2} & C_{3}^{3}  \\
        \end{pmatrix}
        =
        \begin{pmatrix}
            0 & A_{1}^{2} & A_{1}^{3} \\
            \grey{-A_{1}^{2}} & 0 & A_{2}^{3} \\
            \grey{-A_{1}^{3}} & \grey{-A_{2}^{3}} & 0 \\
        \end{pmatrix}
        \blue{
        \begin{pmatrix}
            B_{1}^{1} & B_{1}^{2} & B_{1}^{3} \\
            B_{2}^{1} & B_{2}^{2} & B_{2}^{3} \\
            B_{3}^{1} & B_{3}^{2} & B_{3}^{3} 
        \end{pmatrix}
        }
    \end{equation*}
    }

    Output
    \small{
    \setlength{\jot}{2pt} 
    \begin{equation*}
        \begin{aligned}
            C_{1}^{1} &= 2m_{1} - m_{2} - m_{3} + 2m_{4} + m_{5} + m_{6} - 2m_{8} + m_{9} \\ 
                      &\phantom{=}- 2m_{10} - 2m_{11} + 2m_{13} - 2m_{14} \\
            C_{1}^{2} &= m_{1} - m_{2} + m_{5} - m_{8} + m_{9} - 2m_{10} - m_{11} + 2m_{13} \\
            C_{1}^{3} &= m_{1} - m_{3} + m_{4} + m_{6} - m_{8} - 2m_{10} - m_{11} + 2m_{13} - m_{14} \\
            C_{2}^{1} &= m_{1} - m_{2} - m_{3} + m_{4} - 2m_{7} - m_{8} - m_{10} - m_{11} \\
                      &\phantom{=}+ m_{12} + m_{13} - m_{14} \\
            C_{2}^{2} &= m_{1} - m_{2} - 2m_{7} - m_{8} - m_{10} + m_{12} + m_{13} \\
            C_{2}^{3} &= m_{1} - m_{2} - m_{3} + 2m_{4} - m_{5} - 2m_{7} - m_{8} - m_{9} \\
                      &\phantom{=}- m_{10} + m_{12} + m_{13} - m_{14} \\
            C_{3}^{1} &= m_{2} - m_{3} + m_{5} - m_{6} - 2m_{7} - m_{9} + 2m_{12} \\
            C_{3}^{2} &= -m_{1} + m_{2} + m_{5} - m_{8} - m_{9} - m_{11} \\
            C_{3}^{3} &= -m_{7} - m_{10} + m_{12}
        \end{aligned}
    \end{equation*}
    }

\end{minipage}

\vspace{-0.2cm}
\caption{
    \justifying
    $\langle 3, 3, 3 \colon 14 \rangle_\stmm{kg}^{(0,0,0)}$. Batched cross product.
    Coefficients lie in \(\mathbb{Q}\). 
    Operation count: 14 multiplications and 126 additions.
}
\label{fig:kg-333}
\end{figure*}

\begin{figure*}

\begin{minipage}[t]{0.39\textwidth}
    \raggedright
    General multiplications
    \small{
    \begin{equation*}
        \begin{aligned}
            m_{1}  &= (A_1^2)\blue{(B_2^1)} \\
            m_{2}  &= (A_2^4)\blue{(B_4^4)} \\
            m_{3}  &= (A_1^3 + \red{A_3^3})\blue{(B_3^1)} \\
            m_{4}  &= (A_1^2)\blue{(B_2^2 + B_3^2)} \\
            m_{5}  &= (A_2^2 - A_2^3)\blue{(B_3^4)} \\
            m_{6}  &= (A_1^2 - A_1^3 - A_1^4)\blue{(B_3^2)} \\
            m_{7}  &= (A_2^4 - A_3^4)\blue{(B_3^3 - B_4^3)} \\
            m_{8}  &= (\red{A_1^1} - A_1^4 - A_3^4)\blue{(B_4^1)} \\
            m_{9}  &= (\red{A_3^3} + A_3^4)\blue{(B_4^3 - B_4^4)} \\
            m_{10} &= (\red{A_3^3} + A_3^4)\blue{(B_4^1 - B_4^2)} \\
            m_{11} &= (\red{A_1^1} - A_1^4)\blue{(B_3^2 - B_4^2)} \\
            m_{12} &= (A_2^3 + A_2^4 - \red{A_3^3} - A_3^4)\blue{(B_3^3)} \\
            m_{13} &= (\red{A_1^1} - A_1^4 + \red{A_3^3})\blue{(B_3^2 + B_4^1 - B_4^2)} \\
            m_{14} &= (A_1^3 - \red{A_2^2} + A_2^3)\blue{(B_2^1 - B_2^4 - B_3^4)} \\
            m_{15} &= (A_2^4 - \red{A_3^3} - A_3^4)\blue{(B_3^3 - B_4^3 + B_4^4)} \\
            m_{16} &= (A_1^3 + A_2^3 + A_2^4 - A_3^4) \\[-1.5mm]&\hspace{4.2mm} \blue{(B_3^3 - B_4^1 - B_4^3)} \\
            m_{17} &= (A_1^2 - A_1^3 - A_1^4 - A_2^3 - A_2^4) \\[-1.5mm]&\hspace{4.2mm} \blue{(B_2^2 - B_2^3 + B_3^2)} \\
            m_{18} &= (A_1^2 - A_1^3 - A_1^4 + \red{A_2^2} - A_2^3) \\[-1.5mm]&\hspace{4.2mm} \blue{(B_3^2 - B_4^2 + B_4^4)} \\
            m_{19} &= (A_1^3 + A_2^3) \\[-1.5mm]&\hspace{4.2mm} \blue{(B_2^1 - B_2^4 + B_3^1 + B_3^3 - B_3^4 - B_4^1 - B_4^3)} \\
            m_{20} &= (A_1^2 - A_1^3 + \red{A_2^2} - A_2^3) \\[-1.5mm]&\hspace{4.2mm} \blue{(B_2^1 - B_2^4 - B_3^2 + B_4^2 - B_4^4)} \\
            m_{21} &= (A_1^3 + A_1^4 + A_2^3 + A_2^4) \\[-1.5mm]&\hspace{4.2mm} \blue{(B_2^2 - B_2^3 + B_3^2 - B_4^1 - B_4^3)} \\
            m_{22} &= (A_1^2 - A_1^3 - A_1^4 + \red{A_2^2} - A_2^3 - A_2^4) \\[-1.5mm]&\hspace{4.2mm} \blue{(B_2^2 - B_2^3 + B_3^2 - B_4^2 + B_4^4)} \\
        \end{aligned}
    \end{equation*}
    }
\end{minipage}
\hfill
\begin{minipage}[t]{0.6\textwidth}
    \raggedright
    Block layout
    \small{
    \begin{equation*}
        \begin{pmatrix}
            C_{1}^{1} & C_{1}^{2} & C_{1}^{3} & C_{1}^{4} \\
            C_{2}^{1} & C_{2}^{2} & C_{2}^{3} & C_{2}^{4} \\
            C_{3}^{1} & C_{3}^{2} & C_{3}^{3} & C_{3}^{4} \\
            C_{4}^{1} & C_{4}^{2} & C_{4}^{3} & C_{4}^{4}
        \end{pmatrix}
        =
        \begin{pmatrix}
            \red{A_{1}^{1}} & A_{1}^{2} & A_{1}^{3} & A_{1}^{4} \\
            0 & \red{A_{2}^{2}} & A_{2}^{3} & A_{2}^{4} \\
            0 & 0 & \red{A_{3}^{3}} & A_{3}^{4} \\
            0 & 0 & 0 & \red{A_{4}^{4}}
        \end{pmatrix}
        \;
        \blue{
        \begin{pmatrix}
            B_{1}^{1} & B_{1}^{2} & B_{1}^{3} & B_{1}^{4} \\
            B_{2}^{1} & B_{2}^{2} & B_{2}^{3} & B_{2}^{4} \\
            B_{3}^{1} & B_{3}^{2} & B_{3}^{3} & B_{3}^{4} \\
            B_{4}^{1} & B_{4}^{2} & B_{4}^{3} & B_{4}^{4}
        \end{pmatrix}
        }
    \end{equation*}
    }

    \phantom{42}

    Recursive multiplications
    \small{
    \begin{equation*}
        \begin{aligned}
            \reccall{r_{1}}  &= (\red{A_2^2})\blue{(B_2^3)} \\
            \reccall{r_{2}}  &= (\red{A_4^4})\blue{(B_4^3)} \\
            \reccall{r_{3}}  &= (\red{A_4^4})\blue{(B_4^1)} \\
            \reccall{r_{4}}  &= (\red{A_4^4})\blue{(B_4^2)} \\
            \reccall{r_{5}}  &= (\red{A_1^1})\blue{(B_1^3 - B_4^1)} \\
            \reccall{r_{6}}  &= (\red{A_4^4})\blue{(B_4^1 - B_4^4)} \\
        \end{aligned}
        \hspace{1cm}
        \begin{aligned}
            \reccall{r_{7}}  &= (\red{A_1^1})\blue{(B_1^1 + B_4^1)} \\
            \reccall{r_{8}}  &= (\red{A_1^1})\blue{(B_1^2 - B_1^4)} \\
            \reccall{r_{9}}  &= (\red{A_2^2})\blue{(B_2^4 + B_3^4)} \\
            \reccall{r_{10}} &= (\red{A_1^1})\blue{(B_1^4 - B_3^2 + B_4^2)} \\
            \reccall{r_{11}} &= (\red{A_3^3})\blue{(B_3^1 - B_3^2 - B_4^1 + B_4^2)} \\
            \reccall{r_{12}} &= (\red{A_3^3})\blue{(B_3^3 - B_3^4 - B_4^3 + B_4^4)} \\
        \end{aligned}
    \end{equation*}
    }

    \phantom{42}
    
    Output
    \small{
    \begin{equation*}
        \begin{aligned}
            C_1^1  &= \reccall{r_{7} - r_{11}} + m_{1} + m_{3} + m_{11} - m_{13} \\
            C_1^2  &= \reccall{r_{8} + r_{10}} + m_{4} - m_{6} + m_{11} \\
            C_1^3  &= \reccall{r_{5}} - m_{2} + m_{4} - m_{7} + m_{8} - m_{9} - m_{12} + m_{15} + m_{16} - m_{17} - m_{21} \\
            C_1^4  &= \reccall{r_{10}} + m_{1} + m_{5} + m_{11} - m_{14} - m_{18} - m_{20} \\
            C_2^1  &= \reccall{r_{9} + r_{11}} - m_{3} + m_{7} - m_{8} - m_{11} + m_{13} - m_{14} - m_{16} + m_{19} \\
            C_2^2  &= \reccall{r_{1}} + m_{2} + m_{6} - m_{17} - m_{18} + m_{22}  \\
            C_2^3  &= \reccall{r_{1}} + m_{2} + m_{9} + m_{12} - m_{15} \\
            C_2^4  &= \reccall{r_{9}} + m_{2} - m_{5} \\
            C_3^1  &= \reccall{r_{11}} - m_{8} - m_{11} + m_{13} 
            \hspace{2.3cm}
            C_4^1  = \reccall{r_{3}}
            \\
            C_3^2  &= -m_{8} - m_{10} - m_{11} + m_{13} 
            \hspace{2cm}
            C_4^2  = \reccall{r_{4}} 
            \\
            C_3^3  &= m_{2} + m_{7} + m_{9} - m_{15} 
            \hspace{2.52cm}
            C_4^3  = \reccall{r_{2}} 
            \\
            C_3^4  &= \reccall{-r_{12}} + m_{2} + m_{7} - m_{15} 
            \hspace{2.21cm}
            C_4^4  = \reccall{r_{3} - r_{6}} 
            \\
        \end{aligned}
    \end{equation*}
    }
\end{minipage}

\caption{
    \justifying
    $\langle 4, 4, 4 \colon 34 \rangle_\stmm{ug}^{(12,0,0)}$.
    Structured matrix multiplication \(C = A B\) for a \(4\times4\) block with upper-triangular \(A\).
    Coefficients lie in \(\mathbb{Z}\).
    % ($\{-1,0,1\}$)
    Operation count: 34 multiplications ($12\, \stmm{ug} + 22\,\stmm{gg}$) and 148 additions.
}

\label{fig:ug-444}
\end{figure*}

\begin{figure*}

\begin{minipage}[t]{0.46\textwidth}
    \raggedright
    General multiplications
    \small{
    \begin{equation*}
        \begin{aligned}
            m_{1} &= (A_{1}^{2} - A_{2}^{2})\blue{(B_{1}^{2} - B_{2}^{2})} \\
            m_{2} &= (A_{1}^{1} + A_{1}^{2})\blue{(B_{1}^{1} + B_{1}^{2})}\inverseoftwo{/2} \\
            m_{3} &= (A_{1}^{1} - A_{2}^{2})\blue{(B_{1}^{2})} \\
            m_{4} &= (A_{1}^{1} - A_{1}^{2})\blue{(B_{1}^{1} - B_{1}^{2})}\inverseoftwo{/2} \\
            m_{5} &= (A_{1}^{2})\blue{(B_{1}^{1} - B_{2}^{2})}
        \end{aligned}
    \end{equation*}
    }
\end{minipage}
\hfill
\begin{minipage}[t]{0.5\textwidth}
    \raggedright
    Block layout
    \small{
    \begin{equation*}
        \begin{pmatrix}
            C_{1}^{1} & C_{1}^{2} \\
            C_{2}^{1} & C_{2}^{2} \\
        \end{pmatrix}
        =
        \begin{pmatrix}
            A_{1}^{1} & A_{1}^{2} \\
            \grey{A_{1}^{2}} & A_{2}^{2}
        \end{pmatrix}
        \blue{
        \begin{pmatrix}
            B_{1}^{1} & B_{1}^{2} \\
            \grey{B_{1}^{2}} & B_{2}^{2} 
        \end{pmatrix}
        }
    \end{equation*}
    }

    % \phantom{42}

    Output
    \small{
    \begin{equation*}
        \begin{aligned}
            C_{1}^{1} &= m_2 + m_4 \\
            C_{1}^{2} &= m_2 - m_4 - m_5 \\
            C_{2}^{1} &= m_2 - m_3 - m_4 \\
            C_{2}^{2} &= m_1 + m_2 - m_3 - m_4 - m_5
        \end{aligned}
    \end{equation*}
    }

\end{minipage}

\caption{
    \justifying
    $\langle 2, 2, 2 \colon 5 \rangle_\stmm{ss}^{(0,2,0)}$.
    Coefficients lie in \(\mathbb{Q}\). 
    Operation count: 5 multiplications and 17 additions.
}
\label{fig:ss-222}
\end{figure*}

\clearpage

\section{Appendix}

\subsection{Derivation of the asymptotic complexity ratio} \label{app:gamma-ab}

We analyze the recurrence governing recursive structured matrix multiplication schemes. Let $k \geq 2$ be the block size, $r$ the total rank (number of scalar products) of the base scheme, and $\omega$ the matrix multiplication exponent used at recursion (so $M_{\stmm{gg}}(n)=n^\omega$). Denote by $q_{\stmm{ab}}, q_{\stmm{ag}}, q_{\stmm{gb}}$ the numbers of recursive calls that preserve, respectively, both structures $\stmm{ab}$, only the left structure $\stmm{ag}$, and only the right structure $\stmm{gb}$. At size $n$, the master recurrence reads
\begin{equation*}
\begin{aligned}
    M_{\stmm{ab}}(n)
    &= q_{\stmm{ab}}\,M_{\stmm{ab}}(\tfrac{n}{k})
    % \\ &\phantom{=}
    + q_{\stmm{ag}}\,M_{\stmm{ag}}(\tfrac{n}{k})
    + q_{\stmm{gb}}\,M_{\stmm{gb}}(\tfrac{n}{k})
    \\ &\phantom{=}
    + (r - q_{\stmm{ab}} - q_{\stmm{ag}} - q_{\stmm{gb}})\,M_{\stmm{gg}}(\tfrac{n}{k}).
\end{aligned}
\end{equation*}
Throughout, we first assume $n=k^m$ for some integer $m\ge 0$; the usual padding/smoothing argument implies the same asymptotics for arbitrary $n$.

Assume $q_{\stmm{ag}}=q_{\stmm{gb}}=0$. Then recurrence simplifies to
\begin{equation*}
    M_{\stmm{ab}}(n) = q_{\stmm{ab}}\,M_{\stmm{ab}}(n/k) + \bigl(r-q_{\stmm{ab}}\bigr)\,(n/k)^\omega.
\end{equation*}
Setting $n=k^m$, assuming $M_{\stmm{ab}}(1)=1$, unrolling it gives
\begin{align*}
    M_{\stmm{ab}}(k^m)
    &= q_{\stmm{ab}}^{\,m}
    + \bigl(r-q_{\stmm{ab}}\bigr)\sum_{j=0}^{m-1} q_{\stmm{ab}}^{\,j}\,k^{\omega(m-1-j)} \nonumber\\
    &= q_{\stmm{ab}}^{\,m}
    + \bigl(r-q_{\stmm{ab}}\bigr)\,\frac{k^{\omega m}-q_{\stmm{ab}}^{\,m}}{k^\omega-q_{\stmm{ab}}}.
\end{align*}
Rewriting in terms of $n=k^m$ yields the exact closed form
\begin{equation*}
    M_{\stmm{ab}}(n)
    = n^{\log_k q_{\stmm{ab}}}
    + \frac{r-q_{\stmm{ab}}}{k^\omega-q_{\stmm{ab}}}\,\bigl(n^\omega - n^{\log_k q_{\stmm{ab}}}\bigr).
\end{equation*}
When $k^\omega>q_{\stmm{ab}}$, the leading term is $n^\omega$ and hence
\begin{equation*}
    \gamma_{\stmm{ab}} \overset{\mathrm{def}}{=} \lim_{n\to\infty}\frac{M_{\stmm{ab}}(n)}{M_{\stmm{gg}}(n)} = \frac{r-q_{\stmm{ab}}}{k^\omega-q_{\stmm{ab}}}.
\end{equation*}
For the more general case $q_{\stmm{ag}} \neq 0$ and $q_{\stmm{gb}}\neq0$, assume we already have asymptotics
\begin{align*}
    M_{\stmm{ag}}(n) &= \gamma_{\stmm{ag}}\,n^\omega + o(n^\omega), \\
    M_{\stmm{gb}}(n) &= \gamma_{\stmm{gb}}\,n^\omega + o(n^\omega).
\end{align*}
We seek a solution of the form $M_{\stmm{ab}}(n) = \gamma_{\stmm{ab}}\,n^\omega + o(n^\omega)$. Substituting asymptotics into the master recurrence gives
\begin{equation*}
\gamma_{\stmm{ab}}
= \frac{r - q_{\stmm{ab}} - q_{\stmm{ag}}(1 - \gamma_{\stmm{ag}}) - q_{\stmm{gb}}(1 - \gamma_{\stmm{gb}})}
       {k^\omega - q_{\stmm{ab}}}.
\end{equation*}

\subsection{Block-Recursive Baseline} \label{app:baseline}
% \paragraph{Baseline from matrix--vector building blocks.}
When a dedicated structured matrix--matrix algorithm is unavailable (essentially, all structures except $\stmm{gt}$), we adopt a conservative baseline that assembles $C=AB$ from two ingredients: (i) structural zeros; and (ii) the best available structured matrix--vector routines. Concretely, for a chosen block size $k \geq 2$ we partition $A$ and $B$ into $k\times k$ blocks and compute $C$ columnwise: each block column of $B$ is multiplied by $A$ using the structure-specific matrix--vector algorithm, while structural zeros prune calls that would otherwise arise. This construction induces a recurrence with explicit counts of recursive subproblems; the resulting leading constant $\gamma$ then follows directly from \eqref{eq:factor}. 

As an illustrative example, consider $A$ upper triangular and $B$ general ($\stmm{ug}$) with $k=2$. 
\begin{equation*}
    % \begin{pmatrix} c_1 & c_2\\ c_3 & c_4 \end{pmatrix} =
    \begin{pmatrix} \red{a_1} & a_2\\ 0 & \red{a_3} \end{pmatrix}
    \begin{pmatrix} b_1 & b_2\\ b_3 & b_4 \end{pmatrix}
    = 
    \begin{pmatrix} \red{a_1} b_1 + a_2 b_3 & \red{a_1} b_2 + a_2 b_4 \\ \red{a_3} b_3 & \red{a_3} b_4 \end{pmatrix}
\end{equation*}
There are four recursive calls that preserve the $\stmm{ug}$ structure and two $\stmm{gg}$ calls. Plugging these counts into \eqref{eq:factor} with $\omega=\log_2 7$ gives
\begin{equation*}
    \gamma_{\stmm{ug}}
    = \frac{r - q_{\stmm{ug}}}{2^\omega - q_{\stmm{ug}}}
    = \frac{6 - 4}{7 - 4}
    = \frac{2}{3}.
\end{equation*}
For a general $k\times k$ blocking of the same $\stmm{ug}$ case, each of the $k^2$ output blocks receives exactly one contribution from each diagonal block of $A$, hence $q_{\stmm{ug}}=k^2$. The strictly upper blocks of $A$ contribute $\stmm{gg}$ products: there are $k$ block columns, and in column $i$ there are $(k-i)$ such blocks, for a total of $\sum_{i=1}^{k}(k-i)=\tfrac{k(k-1)}{2}$ per block row of $B$, i.e., $k\cdot \tfrac{k(k-1)}{2}=\tfrac{k^2(k-1)}{2}$ calls. Substituting these counts into \eqref{eq:factor} yields the baseline prediction
\begin{equation*}
    \gamma_{\stmm{ug}}(k)
    = \frac{\tfrac{1}{2}k^2(k-1)}{k^\omega - k^2}.
\end{equation*}
Since $\omega>2$, we have $\gamma_{\stmm{ug}}(k)\sim \tfrac{1}{2}k^{3-\omega}$, which increases with $k$; thus the minimum is attained at $k=2$, giving $\gamma_{\stmm{ug}}=\tfrac{2}{3} \approx 0.667$.

Recent work ~\cite{rybin_2025_10}, developed in the context of exact causal attention, provides a \(4\times4\) scheme that can be interpreted in our notation as \(\langle 4,4,4:34\rangle^{(10,0,0)}_{\stmm{ug}}\). Substituting these parameters into~\eqref{eq:factor} yields
\[
\gamma_{\stmm{ug}}^{\text{ECA}}
= \frac{34 - 10}{4^{\omega} - 10},
\]
so that \(\gamma_{\stmm{ug}}^{\text{ECA}} = 4/9 \approx 0.444\) for \(\omega=3\) and \(\gamma_{\stmm{ug}}^{\text{ECA}} = 8/13 \approx 0.615\) for \(\omega = \log_2 7\). These are the baseline values reported for \(\stmm{ug}\) in Table~\ref{tab:relcomplex}.

\subsection{Implementation Notes} \label{app:implementation}
The implementation is self-contained with no external dependencies, 
consisting of approximately 4000 lines of C++ code. All arithmetic 
operations over finite fields are performed using bitwise-parallel 
techniques on 64-bit words.

For $\F_2$, each vector of field elements is packed into a single 
\texttt{uint64}, enabling bitwise XOR for addition and standard 
population count for inner products. For $\F_3$, we use a two-bit 
encoding: $0 \mapsto 00$, $1 \mapsto 01$, $2 \mapsto 10$, stored 
as a pair of \texttt{uint64} words $(l, h)$ holding low and high 
bits respectively. Addition and negation reduce to a small number 
of bitwise operations on these pairs. This packed representation 
limits the implementation to $n \leqslant 8$, which suffices for 
the base sizes considered in this work.

Hensel lifting from $\F_p$ to $\F_{p^k}$ (typically $k = 10$) reduces to repeatedly solving linear systems modulo $p$. The resulting $p$-adic approximation is then converted to rational coefficients via standard rational reconstruction. We precompute a single row echelon factorization of the Jacobian matrix over $\F_p$, then reuse it across all lifting steps. The bit-packed representation enables efficient Gaussian elimination: each row operation processes 64 field elements in parallel, and the factorization is performed once per scheme rather than once per lifting iteration.

\end{document}